\documentclass[onecolumn,11pt]{article}

\usepackage{chemformula} 
\usepackage[T1]{fontenc} 

\usepackage{mathtools} 
\usepackage{comment}

\usepackage{siunitx}

\usepackage{graphicx}

\usepackage[margin=2.8cm]{geometry}

\usepackage{siunitx}
\DeclareSIUnit{\cal}{cal}
\DeclareSIUnit{\hartree}{\text{\ensuremath {E}}_{\mathrm{h}}}

\usepackage[backend=bibtex,sorting=none,safeinputenc,
,style=numeric-comp
]{biblatex}
\usepackage{hyperref}
\hypersetup{
pdfencoding=auto,
colorlinks=true,
linkcolor=blue!65!red,
menucolor=blue!65!red,
citecolor=green!55!black,
urlcolor=black!50!white!50!blue,
linktocpage=true,
linktoc=all
}

\newcommand*{\etal}{\emph{et al\@.}}
\newcommand*{\Cv}{C_V}
\newcommand*{\ehomo}{E_{\text{HOMO}}}
\newcommand*{\elumo}{E_{\text{LUMO}}}
\newcommand*{\norm}[1]{\lVert #1 \rVert}

\newcommand*{\hpi}[2]{\num[round-mode=places, round-precision=2]{#1},\num[round-mode=places, round-precision=2]{#2}}
\newcommand*{\gpi}[2]{\num[round-mode=places, round-precision=2]{#1},\num[round-mode=places, round-precision=2]{#2}}
\newcommand*{\zpvepi}[2]{\num[round-mode=places, round-precision=4]{#1},\num[round-mode=places, round-precision=4]{#2}}
\newcommand*{\cvpi}[2]{\num[round-mode=places, round-precision=3]{#1},\num[round-mode=places, round-precision=3]{#2}}
\newcommand*{\parseprop}[1]{\num[round-mode=places, round-precision=4]{#1}}

\author{Leon Alday-Toledo$^1$ \and
    Roberto Bernal-Jaquez$^2$ \and
    Saul Zapotecas-Martinez$^3$ \and
    Jose L. Mendoza-Cortes$^4$
}
\date{\small
$^1$Posgrado en Ciencias Naturales e Ingenier\'ia, Universidad Aut\'onoma Metropolitana Unidad Cuajimalpa, Ciudad de M\'exico, M\'exico\\%
$^2$Departamento de Matem\'aticas Aplicadas y Sistemas, Universidad Aut\'onoma Metropolitana Unidad Cuajimalpa, Ciudad de M\'exico, M\'exico\\%
$^3$Coordinaci\'on de Ciencias Computacionales, Instituto Nacional de Astrof\'\i sica \'Optica y Electr\'onica, Tonantzintla, Puebla 72840, M\'exico\\%
$^4$Department of Chemical Engineering \& Materials Science, Michigan State University, East Lansing, Michigan 48824, United States\\%
$^2$\texttt{email:rbernal@cua.uam.mx}
}

\title{Obtaining transferable chemical insight from solving machine-learning classification problems: Thermodynamical properties prediction, atomic composition as good as Coulomb matrix}

\begin{document}

\maketitle
\thispagestyle{empty}

\begin{abstract}
Machine learning (ML) can be used to construct surrogate models for the fast prediction of a property of interest. ML can thus be applied to chemical projects, where the usual experimentation or calculation techniques can take hours or days for just one sample. In this manner, the most promising candidate samples could be extracted from an extensive database and subjected to further in-depth analysis.

Despite their broad applicability, it can be challenging to apply ML methods to a given chemical problem since a multitude of design decisions must be made, such as the molecular descriptor to use or the optimizer to train the model.

Here we present a methodology for the meaningful exploration of a given molecular problem through classification experiments. This conceptually simple methodology results in transferable insight on the selected problem and can be used as a platform from which prediction difficulty is estimated, molecular representations are tested and refined, and more precise or ambitious projects can be undertaken. 
Physicochemical insight can also be obtained.

This methodology is illustrated through the use of multiple molecular descriptors for the prediction of enthalpy, Gibbs' free energy, zero-point vibrational energy, and constant-volume calorific capacity of the molecules from the public database QM9 \cite{Ramakrishnan2014} with \num{133885} organic molecules. A noteworthy result is that for the classification problem we propose, the low-resolution descriptor `atomic composition' \cite{Tchagang2019} can reach a classification rate almost on par with the high-resolution `sorted Coulomb matrix' \cite{Rupp2012,Montavon2012,Hansen2013} ($>90\%$), provided that an appropriate optimizer is used during training.
\end{abstract}

\newpage
{
\hypersetup{linkcolor=black!60!blue}
\tableofcontents
}

\section{Introduction}

Chemical machine learning (ML) seeks to obtain computationally expensive
molecular properties by means of training a predictor with computationally
simpler information. The predictor learns the underlying patterns
in the data and creates a model that connects the inputs with the outputs. If the input properties are
deeply related to the outputs, we expect the predictions to be accurate.

Chemical ML projects have become feasible in recent years given the broad availability of modern computational resources, both in terms of hardware and software. Furthermore, public molecule datasets have emerged, some with cardinalities in the range of hundreds-of-thousands \cite{Ramakrishnan2014} or even near-billion 
and hundreds-of-billions \cite{Blum2009,Ruddigkeit2012}. 

Molecules have to be encoded by means of a \emph{molecular representation} or \emph{descriptor} in order to be compatible with ML models. This is an underlying component of any ML model, with a strong influence on the prediction quality. One project that explores the relationship between descriptor choice and prediction quality is  that of Faber \etal\ \cite{Faber2017}, who tested nine representations paired with six regressors for nine target properties. Their findings are promissory: Several of their models closely follow B3LYP data, and they could potentially outperform it if a dataset with a higher theory level were available. 

This problem can also be approached from the point of view of mathematics, treating minimization of prediction error as an optimization problem. One such example is that of Browning \etal\ 
 \cite{Browning2017}, who opted to select their training set in a guided manner, using genetic algorithms. The resulting training sets can train a NN to a markedly better performance compared to a NN trained with a randomly-selected training set, which is the usual method. This method was tested on, and proved true, for a multitude of molecular properties, including thermodynamical properties such as $H$, $G$, $\Cv$, electronic properties such as $E_{\text{HOMO}}$, $E_{\text{LUMO}}$, $E_{\text{gap}}$, and others such as dipole moment and isotropic polarizability.

One manner to tackle this problem from the chemist's angle is to consider other molecular representations that could supply a predictor with new information and thus improve its performance. However, carrying this endeavor to completion can prove to be deceptively difficult, as the decision of the set of candidate descriptors is just one of the many choices that have to be made before an experiment can be realized. For example, multiple ML models can be used for prediction, a variety of parameter optimization methods can be used for training the model, and numerous programming languages and ML packages can be considered.

In this paper we shall use neural networks (NNs) to model our predictors: they are a popular choice in terms of literature and implementation, and can be used for problems such as regression and classification. Their broad applicability is due to their `universal approximator' character \cite{Cybenko1989,Hornik1989}. Indeed, as stated by Hornik \etal\ \cite{Hornik1989},
\begin{quote}\em
...any lack of success in applications must arise
from inadequate learning, insufficient numbers of
hidden units or the lack of a deterministic
relationship between input and target.
\end{quote}

Note that other ML models or techniques can be applied to chemical problems, such as support-vector regression ($\varepsilon$-SVR), $\nu$-SVR, and random-forest regression. These three were tested by Aldosari \etal\ \cite{Aldosari2021} for the prediction of constant-pressure calorific capacity and entropy of hydrocarbons. A similar endeavor is that of Liu \etal\ \cite{Liu2019}, where $PVT$ properties of pure compounds \ch{H2O}, \ch{CO2}, \ch{H2}, and ternary mixtures of them, are predicted by means of $\nu$-SVR rather than through the development of complex equations of state. This ML approach yields `extremely satisfactory mapping and prediction results'. 
Other models such as convolutional neural networks (CNN) have been previously applied to chemical problems. For example, Kr\"{u}ger \etal\ \cite{Kruger2022} use CNNs for the prediction of the reduction potential of quinones.

Given that we have selected neural networks for this project, additional choices that must be made are the dataset and optimizer to be used in training, the molecular encoding (the inputs), target properties (outputs), the NN architecture (the amount of layers, the amount of neurons in each one, and the activation functions that connect the layers), and the loss function (the error metric to minimize). Experiments can be performed once these settings are chosen.


The findings obtained from these experiments, such as the identification of high-performance representations, elucidate the selected problem through resulting transferable knowledge. Insight is gained on the involved properties, and on possible future experiments that possess increased precision by selecting a stricter error criteria, or increased ambition by using a dataset with a greater amount and variety of molecules, such as the GDB family \cite{Fink2005,Fink2007,Blum2009,Ruddigkeit2012,GDBhost}, the biggest of which contains 166 billion (\num{1.66E11}) molecules. Alternatively, other ML methods could also be tested.

As we shall show, our target properties were enthalpy ($H$), Gibbs' free energy ($G$), zero-point vibrational energy (ZPVE), and constant-volume calorific capacity ($\Cv$). We trained neural networks (NN) with three hidden layers for molecule classification according to the value of each of their properties. Our experiments reveal that the coarse-grained `atomic composition' representation can be sufficient to classify molecules according to the value of its $H$, $G$, or ZPVE, whereas $\Cv$ prediction is significantly more difficult. Additionally, AdaBelief \cite{AdaBelief-paper} is shown to be a noticeably better optimizer choice than Adam \cite{Adam-paper} for the training of low-resolution models.

Our dataset of choice is QM9 \cite{Ramakrishnan2014}, a popular and public resource constituted of \num{133885} organic molecules with up to nine heavy atoms (\ch{C}, \ch{N}, \ch{O}, \ch{F}). The molecules are presented as simple text files containing the XYZ coordinates of the optimized structure,  plus additional information such as its harmonic frequencies, dipole moment norm, $\ehomo$, $\elumo$, and thermodynamical properties such as zero-point vibrational energy (ZPVE), constant-volume caloriric capacity $\Cv$, enthalpy, free energy, and internal energy. Geometry optimization and property calculations were performed at the level B3LYP/6-31G(2df,p). We opted for using this dataset given its popularity and the wealth of properties it provides.

An alternative dataset choice is PC9 \cite{Glavatskikh2019,Cauchy2019}. Its molecules also have up to 9 heavy atoms, but they show a greater diversity regarding bond distances and functional groups. PC9 is constituted by \num{99234} molecules, \num{80877} of which are not present in QM9; these include \num{4442} radicals and \num{883} triplets; compare this with QM9: all its molecules are closed-shell and neutral. The entries in the PC9 database contain the molecular geometries, atomic charges, total energies, and those of $\text{HOMO}-1$, HOMO, LUMO, $\text{LUMO}+1$, and were calculated at the theory level B3LYP/6-31G(d).

We used $k$-fold cross-validation with $k = 5$, therefore all error decay plots and classification tables show the average values for five experiments with different training and test sets.


Additionally, we modelled our neural networks using the scientific-computing-oriented \texttt{julia} programming language \cite{Bezanson2017}, and the elegant and extensible machine-learning package \texttt{Flux.jl} \cite{Innes2018flux,Innes2018flux2nd}; other \texttt{julia} packages were also used%
 \cite{Innes2018zygote,Besancon2021,Lin2022,Breloff2022,Holy2013,Revels2016,Lin2013,Holy2012,Kuhn2014}.

\subsection{Choice of molecular representation}

In this paper, we shall test multiple molecular representations and compare and analyze their performance so as to identify their relative performances and possible beneficial augmentations. Three representations we term \emph{standalone}, since these use a sole family of properties as NN input. Two additional representations are \emph{hybrid} or \emph{mixed}, since they are the concatenation of two representations:
\begin{itemize}
\item Two standalone representations are highly traditional: the sorted Coulomb matrix (SCM) and its eigenvalues (EV) \cite{Rupp2012,Montavon2012,Hansen2013}.
\item The third one is the atomic composition (AC) \cite{Tchagang2019}.
\item The two hybrids are the concatenations of SCM and AC, and of EV and AC. We abbreviate them SCMAC and EVAC, respectively.
\end{itemize}

The SCM and SCMAC representations encode the molecular stoichiometry and geometry, and their sizes grow quadratically with respect to the molecule's atom count. Therefore, we consider these \emph{high-resolution} (or fine-grained) representations. On the other hand, the EV and EVAC representations scale linearly with the system's size, and AC has constant size for any particular dataset, regardless of the selected molecule. Therefore, these three representations are considered to be \emph{low-resolution} (or coarse-grained).

\section{Theoretical methods}

\subsection{An in-depth look: atomic composition}

The AC representation lists the amount of atoms in a molecule, considering not the molecule alone but the atom types featured across the entire dataset: Formaldehyde, \ch{H2CO}, is QM9's sixth molecule. Since QM9 includes the atom types $(\ch{H},\ch{C},\ch{N},\ch{O},\ch{F})$, formaldehyde's AC representation within the QM9 context is the vector $(2,1,0,1,0)$.

It can be seen that AC is a low-resolution descriptor and it doesn't represent molecules uniquely, namely isomers. Nevertheless, we consider it to encode chemically important information and to be a representation worth testing for these reasons:
\begin{itemize}
\item Burden and Winkler \cite{Burden1996,Burden1999} have employed an idea similar to AC for prediction of molar refractivity, hydrophobicity and biological activity of drug-like molecules.
\item Tchagang and Vald\'es \cite{Tchagang2019} show that the EV and SCM representations can be augmented by concatenation of the AC representation. They predict atomization energy using Bayesian-regularized NNs, for the molecules in the QM7 database \cite{Rupp2015}. The augmented models reach markedly lower prediction errors: their lowest mean absolute error (MAE) is \SI{3.0}{\kilo\cal\per\mole}, reached by an SCMAC model, whereas the lowest MAE reached by SCM models is \SI{4.4}{\kilo\cal\per\mole}.

Their findings show that augmented or mixed representations are a simple yet effective manner to improve prediction quality.
Therefore, AC is a promising auxiliary representation.

\item a wealth of literature exists regarding group-contribution methods. These seek to approximately calculate certain molecular properties from the structure alone. The first of its kind was proposed by Riedel \cite{Riedel1949} in 1949. His method calculates the critical pressure of organic compounds, based only on two variables: the molecule's mass and an additive quantity $\varphi$ calculated from its structure.

Other group-contribution methods calculate the critical properties of pure compounds \cite{Lydersen1955,Ambrose1978,Ambrose1979,Joback1987,Constantinou1994,Nannoolal2007}, boiling and freezing temperatures \cite{Joback1987,Constantinou1994,Nannoolal2004}, constant-pressure calorific capacity \cite{Benson1958,Joback1987,Aldosari2021}, entropy \cite{Benson1958,Aldosari2021}, enthalpy and Gibbs' free energy of vaporization, fusion, and formation \cite{Benson1958,Joback1987,Constantinou1994}, and liquid viscosity \cite{Joback1987}.

There's also Fredenslund \etal's UNIFAC method \cite{Fredenslund1975}, which approximates activity coefficients for components in binary and ternary nonideal liquid mixtures.

While many of these methods consider detailed molecular descriptions involving the contribution of atoms, bonds, and functional groups, we consider that AC can be tested for exploring the thermodynamical landscape of the QM9 dataset. AC can be considered to be analogous to Benson's law of additivity of atomic properties \cite{Benson1958}, which is the zero-order approximation to the law of additivity of molecular properties.
\end{itemize}

Therefore, one of the objectives of this work is to evaluate the performance of AC and compare it with the established representations EV and SCM. Furthermore, we shall also test the ability of mixed representations involving AC and the Coulomb-matrix-derived representations. Since we study the relation between a molecule's stoichiometry and some of its thermodynamical properties, this work can be considered an ML-informed successor to the aforementioned group-contribution methods.

\subsection{An in-depth look: Coulomb-matrix descriptors}

The sorted Coulomb matrices were calculated using the molecular geometries and the python \cite{van1995python} package \texttt{QML} \cite{QMLpackage}. The eigenvalues of the matrices were then calculated with the package \texttt{numpy} \cite{numpypackage}.

The Coulomb matrix (CM) representation \cite{Rupp2012} is a conceptually simple encoding of the geometry. Its elements are defined
\begin{equation}\label{eq:scm}
C_{ij} = \left\{
\begin{array}{cc}
\dfrac{1}{2} Z_i^{2.4} & \text{ if } i = j \\[0.8em]
\dfrac{Z_i Z_j}{\norm{\vec{R}_i - \vec{R}_j}} & \text{ if } i \neq j \\
\end{array}
\right.
\end{equation}
where $Z_i$ is the nuclear charge of atom $i$, and $\vec{R}_i$ refers to its Cartesian coordinates in bohr \cite{Dral2015}. 
The formula for the diagonal elements was obtained from fitting the potential energies of the free atoms to their nuclear charges; the non-diagonal elements represent the Coulomb repulsion between every pair of atoms \cite{Rupp2012}.

The CM representation is invariant to translation or rotation, but not to atom reordering in the coordinate list. On the other hand, the eigenvalue representation is invariant to these three operations, and, despite its low dimensionality, Rupp \etal\ \cite{Rupp2012} show it can be used to predict atomization energies, outperforming other methods such as bond counting \cite{Benson1965} or semiempirical PM6 \cite{Stewart2007}.

Furthermore, Coulomb matrices can be made invariant to atom reordering by means of sorting its rows and columns \cite{Montavon2012,Hansen2013}: the rows and columns of an unsorted CM may be uniquely sorted by arranging them according to sorting the nondecreasing order of their norms. Thus, the sorted Coulomb matrix representation (SCM) is obtained, which we use in this paper. Since the SCM is a symmetrical matrix, we only need to consider its nonrepeated elements for training.

\subsection{Choice of the model's outputs}

The choice of a model's output is deeply related to the input features and the model's purpose: if the selected inputs are numerous, then an adequate model would need a high amount of hidden neurons and layers. This is exacerbated if the model is tasked with identifying or approximating multiple sample properties, or if the relation between the inputs and outputs is complex.

In the interest of keeping the model's training affordable, and our results simple enough to be analyzed, we shall perform classification experiments, rather than regression ones. In the latter, the model outputs a continuous variable signifying the predicted value of a given property of a sample. A target test-set average prediction error can then be defined, and the models' performance can be compared to it. On the other hand, in a classification experiment, the metric to optimize is the classification rate.

Therefore, we must define the classes for our experiment: we can define non-overlapping intervals such that a molecule is considered to belong to class $i$ if its value of a certain property is within the $i$-th interval. These intervals needn't be equally-sized, they merely need to be significantly populated, since having great disparity in the cardinality of the classes can lead to heterogeneous prediction behavior and estimating the generalization error can become difficult.

The advantages of this classification technique are that both the model's global and per-class performance are evaluated. The latter can reveal the worst-performing regions of the dataset; focusing on them to improve the model's overall performance can be more effective than focusing on improving the global behavior. Furthermore, once our classification model is trained, it can be used to swiftly filter a database and find the subset of molecules whose value of a given property is within a desired range. Afterwards, these top candidates can be subjected to computationally expensive electronic structure calculations in order to narrow down the list of candidate structures even further.

On the other hand, if our experiments were unable to find a well-performing model, some of the obtained knowledge can be transferable to future experiments, such as the classes that show the easiest prediction, which representations among a selection have the best performance and how they could be improved, or which optimizer is the best suited for training models with the selected inputs and outputs. Therefore, this exploration would still inform and support any other future experiments.

Some classification experiments could have classes with widths beyond the typical computational-chemistry benchmarks. Note, however, that as long as the cardinalities of the classes are not significantly heterogeneous, there is no formal restriction in terms of how many or how thin the classes of a given experiment may be. Therefore, subsequent experiments could use greater granularity by increasing the amount of classes (and thus decreasing their width); regression experiments would also be an option.

If we were to seek a regression model that predicts the value of one property, the NN's output layer would have a single neuron containing the predicted value for a particular sample. On the other hand, for our classification experiments we have as many output neurons as there are classes: neuron $i$ contains the degree of similarity between the current sample and class $i$. Therefore, by deciding how many classes or intervals of interest our experiment shall have, we are also deciding the amount of output neurons in the model.

\subsection{Data preparation}

We standardized the values of the inputs and output properties before performing the experiments, as is commonly done in ML experiments so that inputs and outputs with different absolute values can be used in the same model.

The molecular descriptors we selected have shown to be effective for the prediction of thermodynamical properties. Therefore, the output properties we selected are enthalpy ($H$), Gibbs' free energy ($G$), zero-point vibrational energy (ZPVE), and constant-volume calorific capacity ($\Cv$). Some of their distribution properties are shown in table~\ref{tab:CZstats}.

\begin{figure}[h]
\centering
\includegraphics[page=1]{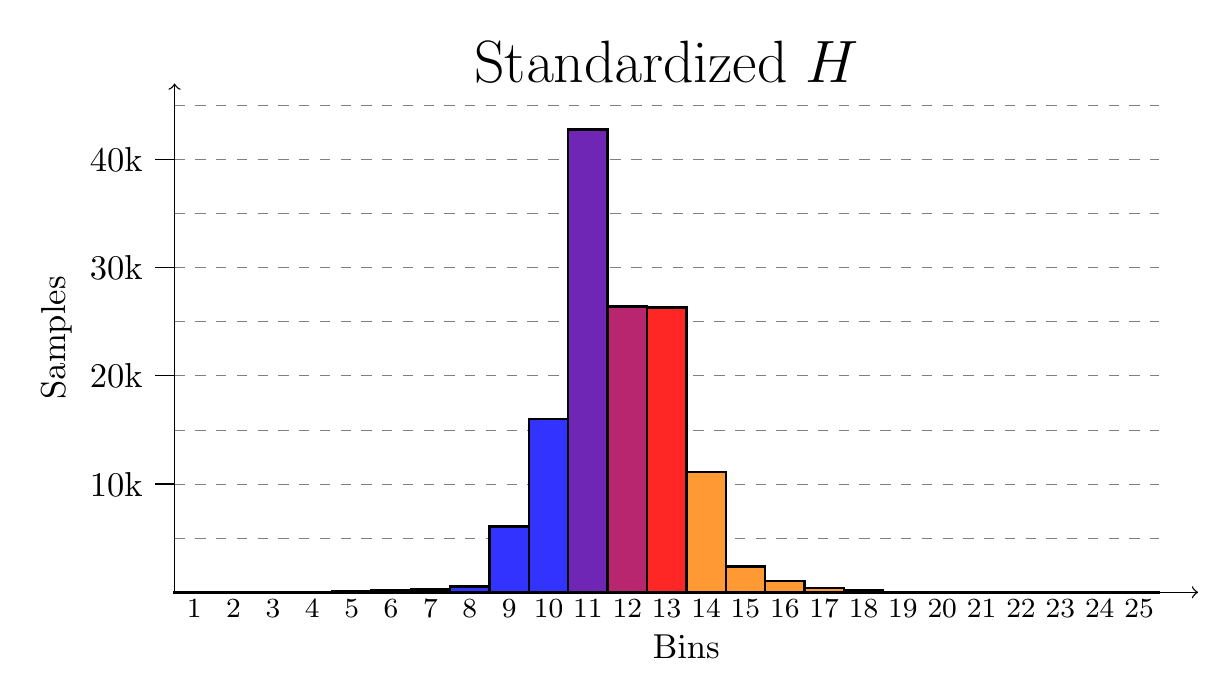}
\caption{Histogram of $H$ and its designated classes.}
\label{fig:histogramH}
\end{figure}

\begin{figure}[h]
\centering
\includegraphics[page=2]{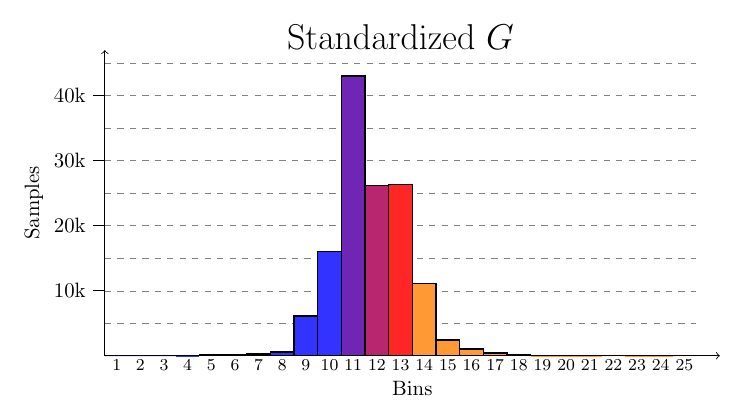}
\caption{Histogram of $G$ and its designated classes.}
\label{fig:histogramG}
\end{figure}

\begin{figure}[h]
\centering
\includegraphics[page=4]{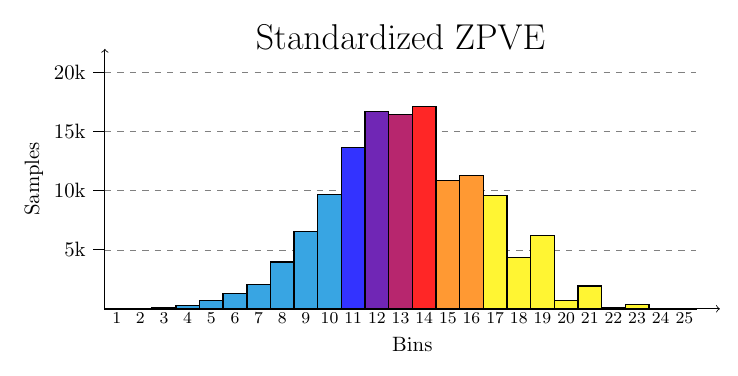}
\caption{Histogram of ZPVE and its designated classes.}
\label{fig:histogramZPVE}
\end{figure}

\begin{figure}[h]
\centering
\includegraphics[page=3]{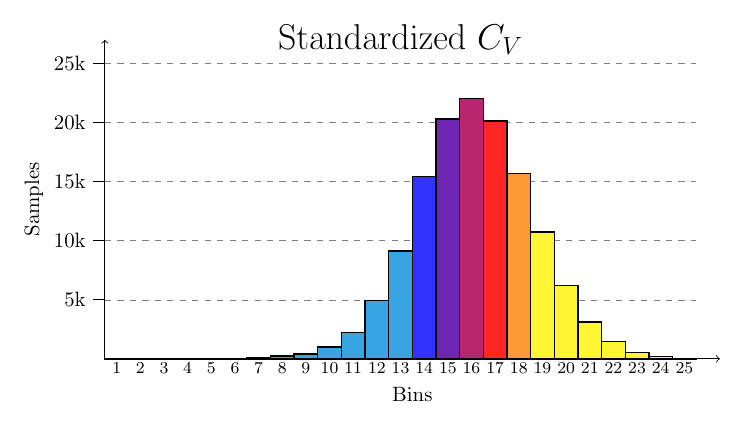}
\caption{Histogram of $\Cv$ and its designated classes.}
\label{fig:histogramCV}
\end{figure}

We must define the amount and thresholds of the classes for each one of these properties. For our experiments, we shall divide the standardized outputs in 25 bins, to be grouped into classes with comparable cardinalities. This leads to $H$ and $G$ having five classes; ZPVE and $\Cv$ having seven; subsequent experiments could use more bins, this leading to more and thinner classes. The current bins and classes are shown in figures~\ref{fig:histogramH} to~\ref{fig:histogramCV}. See also tables~\ref{tab:thresholdsProportionsHG} and~\ref{tab:thresholdsProportionsZPVECV} for their thresholds and populations.

These were the lengths of the input representations we used:
\begin{itemize}
\item AC: as mentioned above, this representation is of length $5$.
\item EV: since the largest QM9 molecules are constituted by $29$ atoms (hydrogens included), our EV representation was fixed at length $29$.
\item SCM: our matrices are of size $29 \times 29$, with a total of $29^2 = 841$ properties. However, the duplicate nondiagonal elements were discarded, and thus this representation had length $\frac{29 \cdot 30}{2} = 435$. For all molecules with less than 29 atoms, their SCMs were padded with rows and columns of zeroes until size $29 \times 29$ was reached, as suggested by the method's authors \cite{Rupp2012,
Montavon2012
}.
\item EVAC and SCMAC: These representations have respective lengths $34$ and $440$. These keep EV and SCM's ability to represent molecules uniquely, unlike AC.
\end{itemize}

\begin{table}[h]
\centering
\begin{tabular}{lS
                 S
                 S
                 S
                 }
\hline
   & {min}    & {max}    & {$\bar{x}$}        & {$\sigma$} \\
\hline
$H$/\si{\hartree}
   & -714.56 & -40.48 & -411.53 & 40.06 \\
$G$/\si{\hartree}
   & -714.60 & -40.49 & -411.58 & 40.06 \\
   ZPVE/\si{\hartree}
   & 0.0160 & 0.274 & 0.149 & 0.0333 \\
$\Cv$/\si{\cal\per\mole\per\kelvin}
   & 6.002    & 46.969   & 31.601 & 4.062  \\
\hline
\end{tabular}
\caption{Statistical parameters of $H$, $G$, ZPVE and $\Cv$ in the QM9 database \cite{Ramakrishnan2014}.}
\label{tab:CZstats}
\end{table}


\begin{table}[h]
\centering
\begin{tabular}{c ccc ccc}
\hline
 & \multicolumn{3}{c}{$H$} & \multicolumn{3}{c}{$G$} \\
\cline{2-4} \cline{5-7}
Class & bins & values/\si{\hartree} & proportion & bins & values/\si{\hartree} & proportion \\
\hline
$c_1$ & $<11$ & $[\hpi{-714.559209}{-444.92557220000003})$ & \parseprop{0.173358} & $<11$ & $[\gpi{-714.602138}{-444.9607216})$ & \parseprop{0.173358} \\
$c_2$ & $11$ & $[\hpi{-444.92557220000003}{-417.96220852})$ & \parseprop{0.319281} & $11$ & $[\gpi{-444.9607216}{-417.99657996})$ & \parseprop{0.321238} \\
$c_3$ & $12$ & $[\hpi{-417.96220852}{-390.99884483999995})$ & \parseprop{0.197289} & $12$ & $[\gpi{-417.99657996}{-391.03243832000004})$ & \parseprop{0.195332} \\
$c_4$ & $13$ & $[\hpi{-390.99884483999995}{-364.03548115999996})$ & \parseprop{0.196624} & $13$ & $[\gpi{-391.03243832000004}{-364.06829668})$ & \parseprop{0.196624} \\
$c_5$ & $>13$ & $[\hpi{-364.03548115999996}{-40.475116999999955})$ & \parseprop{0.113448} & $>13$ & $[\gpi{-364.06829668}{-40.49859700000002})$ & \parseprop{0.113441} \\
\hline
\end{tabular}
\caption{Thresholds and proportions of the classes for $H$ and $G$.}
\label{tab:thresholdsProportionsHG}
\end{table}

\begin{table}[h]
\centering
\begin{tabular}{c ccc ccc}
\hline
 & \multicolumn{3}{c}{ZPVE} & \multicolumn{3}{c}{$\Cv$} \\
\cline{2-4} \cline{5-7}
Class & bins & values/\si{\hartree} & proportion & bins & values/\si{\cal\per\mole\per\kelvin} & proportion \\
\hline
$c_1$ & $<11$ & $[\zpvepi{0.015950999999999993}{0.11914820000000001})$ & \parseprop{0.183949} & $<14$ & $[\cvpi{6.001999999999999}{27.30484})$ & \parseprop{0.135071}\\
$c_2$ & $11$ & $[\zpvepi{0.11914820000000001}{0.12946792000000001})$ & \parseprop{0.102103} & $14$ & $[\cvpi{27.30484}{28.94352})$ & \parseprop{0.115084} \\
$c_3$ & $12$ & $[\zpvepi{0.12946792000000001}{0.13978764000000002})$ & \parseprop{0.124794} & $15$ & $[\cvpi{28.94352}{30.5822})$ & \parseprop{0.151727} \\
$c_4$ & $13$ & $[\zpvepi{0.13978764000000002}{0.15010736000000002})$ & \parseprop{0.122725} & $16$ & $[\cvpi{30.5822}{32.22088})$ & \parseprop{0.164567} \\
$c_5$ & $14$ & $[\zpvepi{0.15010736000000002}{0.16042708})$ & \parseprop{0.127901} & $17$ & $[\cvpi{32.22088}{33.85956})$ & \parseprop{0.150345} \\
$c_6$ & $15,16$ & $[\zpvepi{0.16042708}{0.18106652})$ & \parseprop{0.165276} & $18$ & $[\cvpi{33.85956}{35.49824})$ & \parseprop{0.117018} \\
$c_7$ & $>16$ & $[\zpvepi{0.18106652}{0.2739440000000001}]$ & \parseprop{0.173253} & $>18$ & $[\cvpi{35.49824}{46.969}]$ & \parseprop{0.16618} \\
\hline
\end{tabular}
\caption{Thresholds and proportions of the classes for ZPVE and $\Cv$.}
\label{tab:thresholdsProportionsZPVECV}
\end{table}

\begin{figure}[h]
\centering
\includegraphics{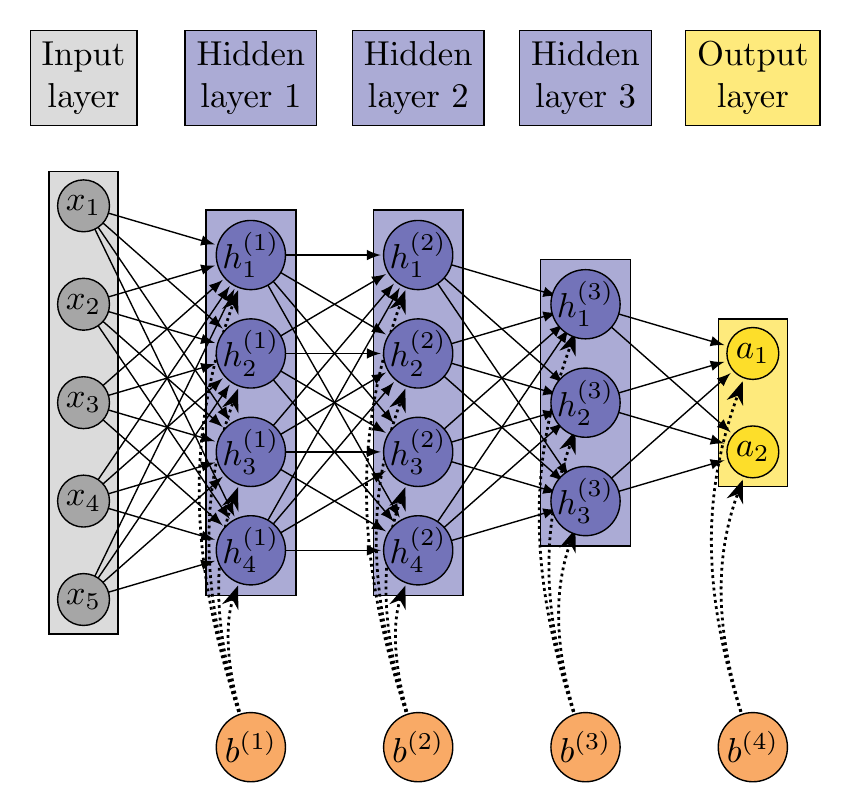}
\caption{Deep neural network (DNN) with three hidden layers.}
\label{fig:nn3-fig}
\end{figure}

\subsection{Underlying machinery of the neural networks}

An illustrative drawing of our NNs can be found in figure~\ref{fig:nn3-fig}. They predict the properties of a sample through a process known as \emph{forward-propagation} \cite{Mitchell1997}, in which the molecule's properties are propagated through the NN's neurons by combination of the inputs with the NN's parameters (its weights and biases), which are processed by the neural network's \emph{activation function} before they reach the next layer.

For classification experiments, the \emph{sigmoid function} (eq.~\ref{eq:sigmoid}) is a common choice. Let $z$ be the model's bias term plus the sum of the product of the model's weights and the sample's properties. Therefore, when $z \to -\infty$, $f(z) \to 0$, and when $z \to \infty$, $f(z) \to 1$.

\begin{equation}\label{eq:sigmoid}
f(z) = \frac{1}{1 + \exp(-z)}
\end{equation}

At the beginning of an ML experiment, the model's parameters are initialized randomly. Since our NNs use the sigmoid activation function, initialization is done through the Xavier initialization scheme\footnote{Originally proposed by Xavier Glorot and Joshua Bengio \cite{xavierglorot}, this method is also known as Glorot initialization.} \cite{xavierglorot}. The output layer's neurons show the similarity between a sample and each of the classes: a perfect model would give a score $1.0$ for the classes the sample belongs to; $0.0$ for those it does not.

Training of the ML model is therefore intended to bring the model's performance as close as possible to the ideal model's: the model's parameters are iteratively adjusted such that the experiment's \emph{loss function} is minimized, which quantifies the model's prediction error for all samples in the training set.

As is typically done in classification experiments, we use the binary cross-entropy loss function:
\begin{equation}\label{eq:bixente}
J_m(W,b) = - \frac{1}{N} \sum_{i = 1}^N \left[ y_i \cdot \ln( a_i ) + (1 - y_i) \ln(1 - a_i) \right]
\end{equation}
where $J_m$ is the error for sample $m$, $N$ is the amount of cells in the output layer (also the amount of classes in the experiment), $y_i$ is the label of sample $m$ for class $i$, and $a_i$ is the model's prediction for class $i$ of sample $m$, a continuous value $\in (0,1)$.

Furthermore, the loss minimization process is performed by means of an optimizer. Many popular choices are gradient-based, such as Adam \cite{Adam-paper,Ruder2016}, a modification of the traditional gradient descent (GD) method. It is an adaptive method since it has a separate step size (\emph{learning rate}) for each parameter, which are reduced at different rates depending on their respective slopes; compare this with GD: it has a fixed, common step size for all parameters even if some parameters have converged and some others have not.

Adam builds upon two gradient-based optimization methods: AdaGrad \cite{AdaGrad-paper} and RMSProp \cite{RMSProp-arxiv-paper}. It possesses the former's ability to deal with sparse gradients and the latter's ability to work with non-stationary problems. We use it for our experiments given its popularity and performance.

We selected a second optimizer: AdaBelief \cite{AdaBelief-paper}, a recent variant of Adam. It seeks to maintain the adaptive methods' stable and fast training, while attaining a test-set (generalization) error similar to that of SGD, which can outperform adaptive methods \cite{Wilson2017}. One of its distinct components is the \emph{belief} that ``the exponential moving average (EMA) of the noisy gradient is a prediction of the gradient at the next time step''. Therefore, it compares the current EMA with the observed gradient for the next step. If they differ, the observed gradient is distrusted and parameters are updated with a smaller step size. Conversely, if they are similar, the effective step size is larger. Given its recency and solid theoretical foundations, we decided to include it our project.

We used the default Adam and Adabelief hyperparameters, just as shown by the authors in their original papers: $\eta = 0.001$, $\beta_1 = 0.9$, $\beta_2 = 0.999$, $\epsilon = \num{1E-8}$.

\subsection{Choice of neural network architecture}

For any given problem, the choice of the amount of layers and their amount of nodes (the model's architecture) is a nontrivial question that requires manual experimentation and testing of multiple candidate architectures; while no certain formulas or procedures exist for the analytical calculation of a perfect architecture, this search can be eased if domain knowledge is available. Another approach to ease the architecture search is through the discipline of `neural network architecture search' (NAS), which encompasses multiple methods that seek to automate this search \cite{Elsken2019,Wistuba2019}.

Some useful guidelines exist, which can be used to obtain candidate architectures intuitively: a neural network should have a funnel-like shape, with early hidden layers having higher neuron counts than the latter layers, and no sharp cardinality drops between one layer and another. Also, the prediction and error performances should be similar for the training and test sets; this indicates the current architecture is experiencing no significant underfit or overfit, and thus can be considered an adequate architecture for the task at hand.

Our input vectors have lengths from $5$ to $440$. Such a wide range warrants using multiple architectures, depending on the amount of features, rather than a single common architecture. Given the aforementioned non-triviality of the neural architecture search, we tested multiple neural network architectures until settling for three-hidden-layer models with the following cardinalities:
\begin{itemize}
\item For AC, \texttt{inp:12:10:8:out},
\item for EV and EVAC, \texttt{inp:26:20:14:out},
\item for SCM and SCMAC, \texttt{inp:240:120:60:out}.
\end{itemize}
where \texttt{inp} is the aforementioned length of the respective feature vector (the standalone representation's lengths are $5$, $29$, and $435$), and \texttt{out} is the amount of classes of the target property ($5$ for $H$ and $G$, $7$ for ZPVE and $\Cv$).

\section{Results and discussion}

\subsection{Enthalpy}
\label{res:enthalpy}

\begin{figure}[h]
\includegraphics[]{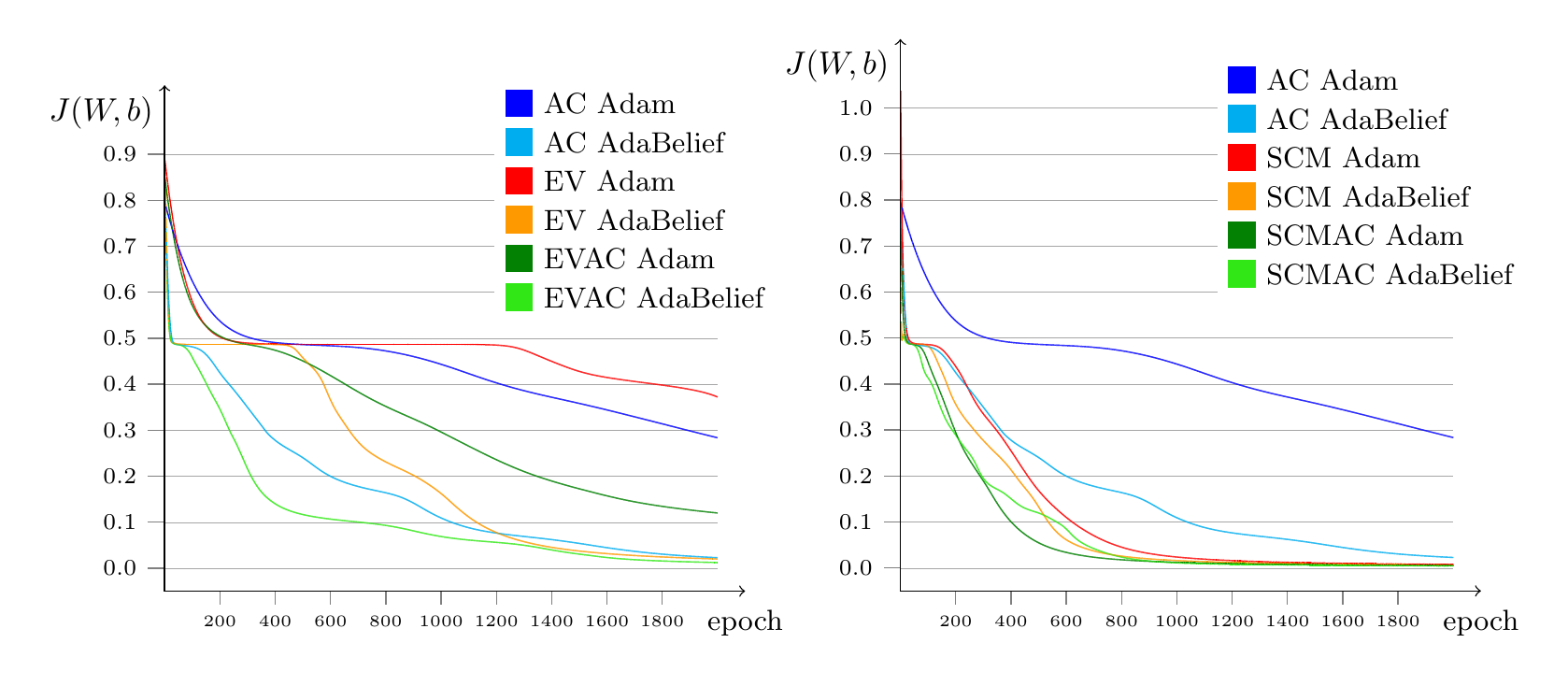}
\caption{Training-set error decay of the enthalpy models.}
\label{fig:h-decays}
\end{figure}

Figure~\ref{fig:h-decays} shows the decay of the training-set error (binary cross-entropy error, equation~\ref{eq:bixente}) after each epoch of training. While a low training error is desirable, a model's ultimate test is the \emph{generalization} or \emph{out-of-sample} error: how would our model perform for heretofore unknown samples, such as those from a future dataset? A simple way to approximate this value is by measuring the model successful classification rate for the molecules in the test set, which didn't take part in the training. This is shown in table~\ref{tab:h-global}. We consider a sample to be classified correctly if the neuron of the sample's class has the biggest score among the output neurons.

As can be seen in figure~\ref{fig:h-decays}, the three AdaBelief curves reach near-zero values, whereas the Adam curves don't. This is consistent with table~\ref{tab:h-global}, where AdaBelief is largely unconcerned by the input features; Adam can reach $>99\%$ classification performances for the high-resolution inputs, while the low-resolution scores are $\in (0.57,0.88)$. Therefore, the optimizer choice is especially important for low-resolution models: the correct optimizer choice can allow a low-resolution model to reach nearly indistinguishable end-point training error and test-set classification performance as a high-resolution model. Furthermore, note that while reaching this performance level requires a training with a greater amount of epochs than that of the high-resolution models, these epochs come at a lower computational cost, given the stark difference in the models' overall amount of neurons.

AC is shown to be a capable standalone descriptor for $H$ prediction. Its influence in the performance of the mixed-representation models is also positive: figure~\ref{fig:h-decays} shows how the mixed-representation models generally train faster than the standalone EV and SCM ones. While the final error of the four models is practically identical, training could be stopped after the cost function has converged, thus the mixed-representation model would spend less computer time than the standalone.

We can draw further noteworthy observations by considering the least-performant experiments:
\begin{itemize}
\item The EVAC representation is the richest out of the low-resolution inputs, given that it includes both EV and AC. Therefore, its performance is the best among its peers.
\item Consider table~\ref{tab:h-global}. Even if we didn't have access to the AdaBelief optimizer and its consistently high performances, we could still perform training experiments with Adam, and improve their performances through the use of richer descriptors.
\item regarding the Adam models, the classification performance of EV and AC are, respectively, $0.5754$ and $0.7279$, compared to EVAC's $0.8783$. We can thus make an argument similar to Hammond's postulate \cite{Hammond1995}, which is usually expressed as the following corollary: \emph{for a given transition state (TS) that connects two minima structures, the TS exhibits greater geometrical similarity to the minimum its energy is closest to}. In our ML context, we can restate it in the following manner: \emph{let A be a standalone representation and M a mixed representation that includes A. Therefore, if A's prediction performance is closest, among its peers, to M's, then A is the most influential representation in M's, among all the representations that constitute M}. This is to say, A is the standalone representation that provides the most valuable information among those in M.

In terms of enthalpy prediction according to our classification experiment, we can thus conclude that AC is a bigger contributor to EVAC's high prediction quality than EV.

This is a simple procedure that can provide us with valuable insight, and we term it \emph{Hammond-like analysis}.
\end{itemize}

%

\begin{table}[h]
\centering
\begin{tabular}{r*{2}{c}}
\hline
\multicolumn{1}{l}{Input} \\
\multicolumn{1}{l}{properties} & Adam & AdaBelief \\
\hline
    AC & 0.7279 & 0.9921 \\
    EV & 0.5754 & 0.9915 \\
  EVAC & 0.8783 & 0.9939 \\
   SCM & 0.9943 & 0.9947 \\
 SCMAC & 0.9948 & 0.9953 \\
\hline
\end{tabular}
\caption{Global classification performance of the enthalpy models for test-set molecules.}
\label{tab:h-global}
\end{table}

\subsection{Gibbs' free energy}
\label{res:gibbs}

This property shows a very similar behavior to enthalpy
. Once again, the high-resolution inputs achieve near-perfect performance; the low-resolution models have an excellent behavior when trained with AdaBelief ($>99\%$ classification performance, table~\ref{tab:g-global}), whereas the Adam-trained models all have scores $\in (0.75,0.94)$.

One noteworthy difference between the $H$ and $G$ models is that while the AC-Adam performance remains steady at $73\%\sim76\%$ for both properties, EV-Adam goes from $0.5754$ for $H$ prediction to $0.8987$ for $G$. This suggests the eigenvalues are more closely related to $G$ than they are to $H$. The EVAC representation is therefore also better suited for $G$ prediction than to $H$; this is visible in tables \ref{tab:h-global} and \ref{tab:g-global}: the Adam scores are, respectively, $0.8783$ and $0.9359$. AdaBelief exhibits minute shifts when comparing $H$ to $G$, but they are consistent with the aforementioned trend.

\begin{figure}[h]
\includegraphics[]{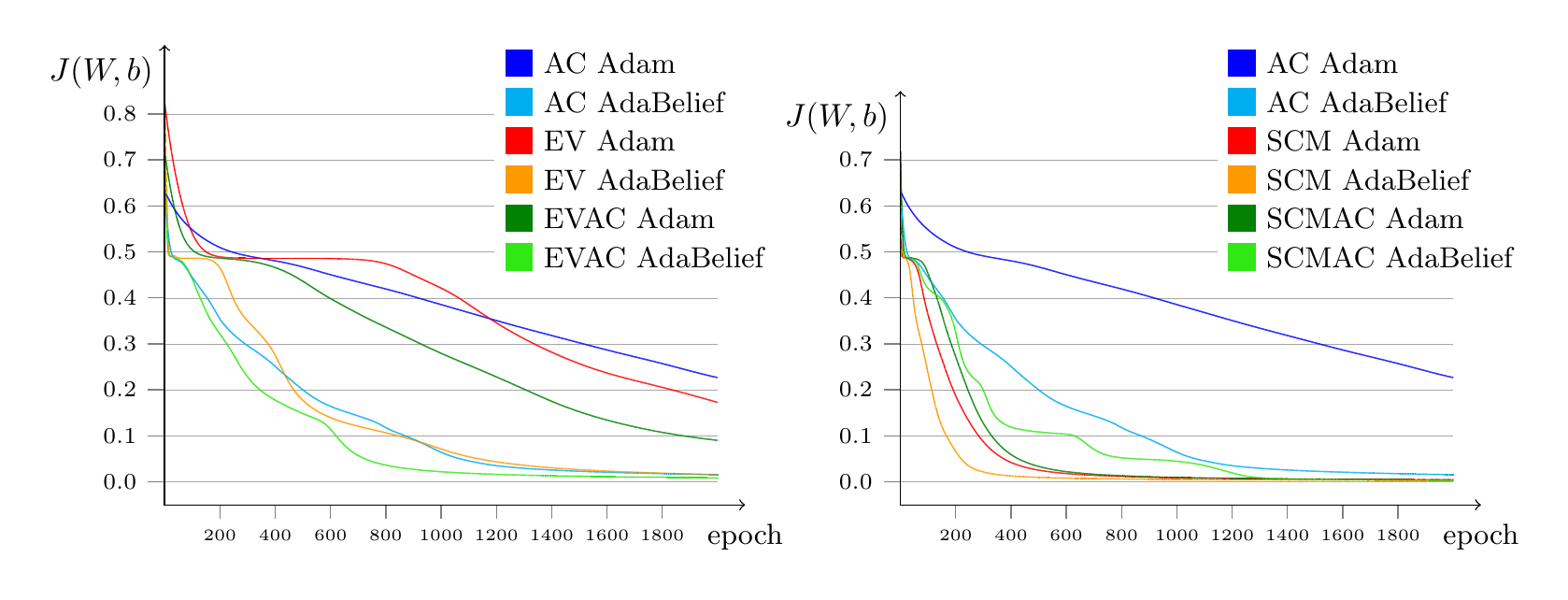}
\caption{Training-set error decay of the Gibbs' free energy models.}
\label{fig:g-decays}
\end{figure}

While the \emph{a priori} table~\ref{tab:thresholdsProportionsHG} and distribution plots~\ref{fig:histogramH} and~\ref{fig:histogramG} show that $H$ and $G$ are nearly identical, the \emph{a posteriori} training curve plots~\ref{fig:h-decays} and~\ref{fig:g-decays} and table~\ref{tab:h-global} and~\ref{tab:g-global} show that these properties are distinct; the Hammond-like analysis shows that EV is more important for $G$ prediction than AC, whereas the converse was true for $H$.

%

\begin{table}[h]
\centering
\begin{tabular}{r*{2}{c}}
\hline
\multicolumn{1}{l}{Input} \\
\multicolumn{1}{l}{properties} & Adam & AdaBelief \\
\hline
    AC & 0.7580 & 0.9913 \\
    EV & 0.8987 & 0.9932 \\
  EVAC & 0.9359 & 0.9952 \\
   SCM & 0.9954 & 0.9955 \\
 SCMAC & 0.9959 & 0.9959 \\
\hline
\end{tabular}
\caption{Global classification performance of the Gibbs' free energy models for test-set molecules.}
\label{tab:g-global}
\end{table}

\subsection{Zero-point vibrational energy}
\label{res:zpve}

This property has a qualitatively similar behavior to that of $H$ and $G$. Once again, the high-resolution descriptors have the best performances, while the low-resolution descriptors perform far better with AdaBelief than with Adam. Furthermore, the Hammond-like analysis suggests that the AC representation is a bigger contributor to the EVAC performance than EV, and it also contains more readily-available relevant information for ZPVE prediction.

\begin{figure}[h]
\includegraphics[]{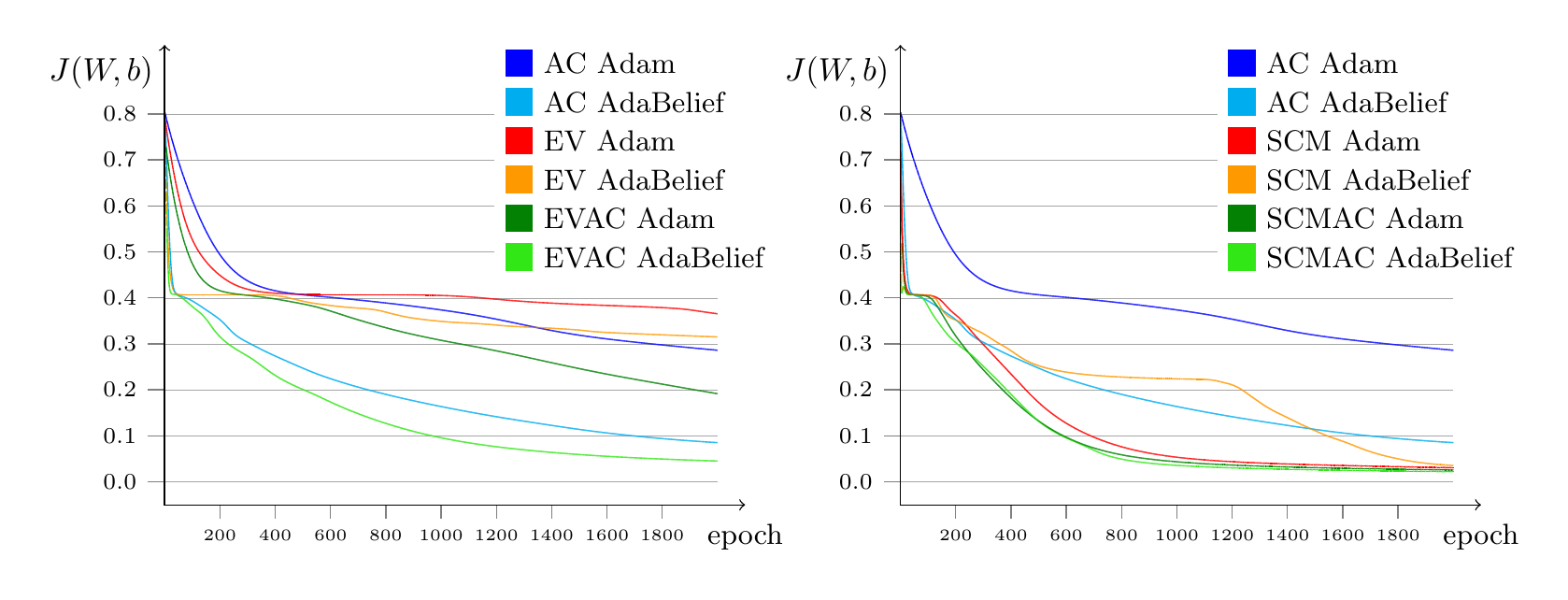}
\caption{Training-set error decay of the ZPVE models.}
\label{fig:zpve-decays}
\end{figure}

Figure~\ref{fig:zpve-decays} shows this property's increased difficulty: the low-resolution models no longer reach near-zero error values; they only reach $<0.1$ on two out of six decays. High-resolution descriptions' performance is also worsened, reaching $0.0221$ 
compared to $H$'s $0.0041$ 
and $G$'s $0.0033$
.

Let us discuss the differences between ZPVE and the two previous properties:
\begin{itemize}
\item AdaBelief reached classification scores $>99\%$ for all $H$ and $G$ models, whereas the ZPVE scores are $\in (0.43,0.97)$. If the EV result is ignored, then the scores are $\in (0.91,0.97)$, which is nonetheless a greater dispersion than that of $H$ and $G$.

\item In order to better understand this property's comparatively high difficulty, table~\ref{tab:zpve-global} includes two additional columns, showing AdaBelief's classification performance for the two classes that exhibit the lowest overall scores.

This per-class analysis shows where our model-refinement efforts may be best employed: rather than seeking a different descriptor with a better global score, we could instead search for a representation that performs adequately for $c_5$ and $c_6$. These additional features could be concatenated to any of the five aforementioned molecular descriptions, and we expect this new mixed representation to show higher $c_5$, $c_6$, and global scores.

EV can be considered such a representation to AC: its $c_6$ performance is better than its dataset average, and its inclusion allows EVAC to nearly rival SCM and SCMAC's performance, both for $c_6$ and globally. On the other hand, EV's contribution to EVAC's $c_5$ performance is much less significant.
\end{itemize}

%

\begin{table}[h]
\centering
\begin{tabular}{r*{4}{c}}
\hline
\multicolumn{1}{l}{Input} & \multicolumn{2}{c}{Global} & \multicolumn{2}{c}{AdaBelief} \\
\cline{2-3} \cline{4-5}
\multicolumn{1}{l}{properties} & Adam & AdaBelief & $c_5$ & $c_6$ \\
\hline
    AC & 0.5069 & 0.9106 & 0.8878 & 0.6686 \\
    EV & 0.3460 & 0.4302 & 0.2830 & 0.4717 \\
  EVAC & 0.7867 & 0.9480 & 0.8959 & 0.8821 \\
   SCM & 0.9573 & 0.9550 & 0.9161 & 0.9070 \\
 SCMAC & 0.9631 & 0.9644 & 0.9307 & 0.9349 \\
\hline
\end{tabular}
\caption{Classification performance of zero-point vibrational energy for test-set molecules.}
\label{tab:zpve-global}
\end{table}

\subsection{Constant-volume heat capacity}
\label{res:cv}

Let us first focus on the similarities between this property and the previous three: AdaBelief outperforms Adam for all models, especially in the low-resolution experiments. Furthermore, the Hammond-like analysis suggests that AC is a better descriptor for $\Cv$ than EV.

\begin{figure}[h]
\includegraphics[]{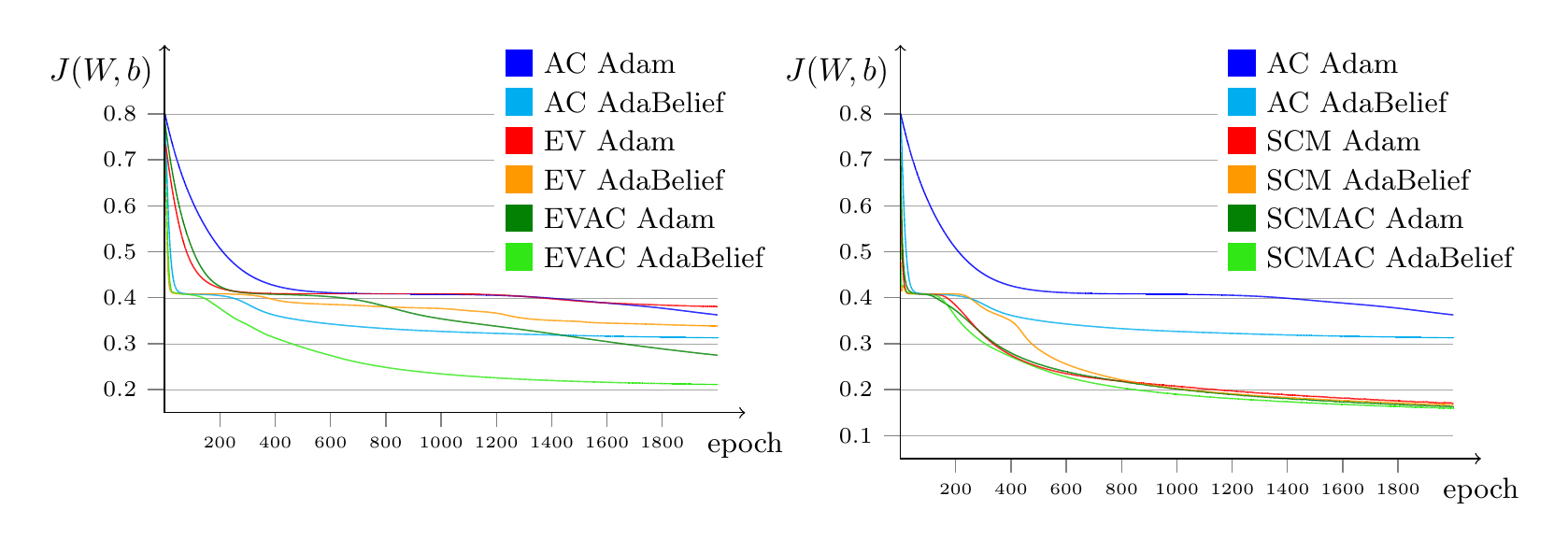}
\caption{Training-set error decay of the $\Cv$ models.}
\label{fig:cv-decays}
\end{figure}

In terms of differences, however, $\Cv$ shows the hardest prediction difficulty. As shown in figure~\ref{fig:cv-decays}, none of the low-resolution models reach errors below $0.2$, whereas the high-resolution models remain $>0.15$ within 2000 epochs. Consequently, the classification scores (table~\ref{tab:cv-global}) are $\in (0.25,0.74)$.

Another difference is visible under AdaBelief: EVAC's classification score is well behind that of the high-resolution models; the previous differences were all $<1\%$. 

In order to shed some light on the unique behavior of $\Cv$, let us turn to table~\ref{tab:cv-adabelief}. This table shows the global and per-class classification performance of AdaBelief alone. This is a more granular display than that of global-only tables, and it provides us with greater insight on the unique behavior of this property.

We can see that AC and EV reach their highest scores for the extrema classes $c_1$ and $c_7$ ($>68\%$), whereas the intermediate classes have scores $<0.5$, some even reaching $<0.1$. That said, the EVAC model shows an improved behavior across all classes; the extrema scores almost reach SCM and SCMAC's, most intermediate classes become $>0.51$, while the exception $c_6$ still shows a marked improvement at $0.4398$. This shows that the molecules of the extrema classes have $\Cv$ values more closely linked to their atomic composition or eigenvalue representation, whereas the molecules of the intermediate classes have other factors influence their $\Cv$.

The low-resolution scores for $c_2$ are also noteworthy: while both standalones have $<0.1$ scores, their mixed representation has performance on par with the other intermediate classes; this sudden performance increase is one of the major reasons behind EVAC's improved global score. Therefore, extrapolating the behavior of mixed representations by comparison of the standalone representations alone can lead to synergistic sets of input features remaining undiscovered. $c_6$ shows a similar behavior.

The performance of any one of the aforementioned representations, low-resolution especially, can be improved significantly by addition of another descriptor with a high score for the intermediate classes. This is especially true for $c_6$, where the gap between low- and high-resolution descriptors is the widest. By improving this class' performance, the cheaper representations would once again approach the more expensive ones.

Alternatively, any of the five representations above can be used to enrich another representation, if its extrema classes' performance is to be improved. In general, a similar procedure can be performed for several candidate representations: once their individual capabilities are found, appropriate high-performance mixed representations can be created. Physical or chemical insight would also be obtained.

%

\begin{table}[h]
\centering
\begin{tabular}{r*{2}{c}}
\hline
\multicolumn{1}{l}{Input} \\
\multicolumn{1}{l}{properties} & Adam & AdaBelief \\
\hline
    AC & 0.3217 & 0.4174 \\
    EV & 0.2548 & 0.3683 \\
  EVAC & 0.5287 & 0.6185 \\
   SCM & 0.7143 & 0.7213 \\
 SCMAC & 0.7285 & 0.7371 \\
\hline
\end{tabular}
\caption{Global classification performance of $\Cv$ for test-set molecules.}
\label{tab:cv-global}
\end{table}

\begin{table}[h]
\centering
\begin{tabular}{r*{8}{c}}
\hline
\multicolumn{1}{l}{Input} & \multicolumn{8}{c}{AdaBelief} \\
\cline{2-9}
\multicolumn{1}{l}{properties} & global  & $c_{1}$ & $c_{2}$ & $c_{3}$ & $c_{4}$ & $c_{5}$ & $c_{6}$ & $c_{7}$ \\
\hline
    AC & 0.4174 & 0.6884 & 0.0879 & 0.4956 & 0.3786 & 0.3423 & 0.0178 & 0.7424 \\
    EV & 0.3683 & 0.7422 & 0.0367 & 0.2732 & 0.2274 & 0.3816 & 0.0000 & 0.7680 \\
  EVAC & 0.6185 & 0.8349 & 0.5670 & 0.5500 & 0.5363 & 0.5156 & 0.4398 & 0.8409 \\
   SCM & 0.7213 & 0.8734 & 0.6768 & 0.6676 & 0.6511 & 0.6489 & 0.6320 & 0.8755 \\
 SCMAC & 0.7371 & 0.8804 & 0.6856 & 0.6871 & 0.6828 & 0.6598 & 0.6374 & 0.8963 \\
\hline
\end{tabular}
\caption{Per-class performance of the constant-volume heat capacity models trained with AdaBelief, for classification of test set molecules.}
\label{tab:cv-adabelief}
\end{table}

\section{Conclusions}

We have tested the performance of 3-hidden-layer NNs for the classification of organic molecules with up to nine heavy atoms, according to the value of their $H$, $G$, ZPVE, and $\Cv$, using coarse-grained experiments.

When low-resolution molecular descriptors are used to train a NN, $>99\%$ classification performance can be reached for $H$ and $G$, both with five classes. ZPVE was divided in 7 classes, and can be classified with $>90\%$ rates. Note that the outcome depends on the optimizer: AdaBelief routinely achieves high classification performance for most descriptors; the popular Adam optimizer is unable to reach such performance for low-resolution inputs.

These results show that, depending on the precision target, low-resolution descriptors with smaller architectures and appropriate optimizers can be sufficient; expensive experiments with high-resolution features and a high neuron- and layer-count may be forgone entirely.

Alternatively, low-resolution chemically-significant representations such as AC may contain highly influential descriptors with new readily-available information that can be extracted from the inputs directly, rather than learned or derived by the NN during training. Therefore, aggregating these descriptors into a mixed representation results in a more descriptive molecular representation, which can lead to improved model performance and reduced training time. This holds true even if no ideal optimizer is known, and can allow one to undertake future experiments seeking more ambitious and chemically diverse datasets or tighter precision targets.

High-resolution representations were also tested. For these experiments, the optimizer choice is much less influential in the outcome: Adam and AdaBelief have very similar performances.

Despite their higher training cost, the performance of the high-resolution models is only slightly ahead of the low-resolution ones, for all but one output property: $\Cv$, which shows the highest prediction difficulty for all optimizers and input descriptors; the best-performant $\Cv$ model uses AdaBelief and the SCMAC representation, reaching a test-set classification score of $0.7371$. Thus, $\Cv$ prediction can still be improved.

As we have shown previously, one way to improve the prediction quality for the more difficult properties is through the creation of an appropriate mixed representation. The per-class and Hammond-like analyses, while simple, can nevertheless provide very valuable information and guide the construction of such a representation. The former can be more revealing than a global-performance analysis in identifying the current descriptor's underperforming classes; the latter can reveal the relative influence of multiple descriptors that take part in a given mixed representation. For example, the Hammond-like analysis on $H$ and $G$ shows that the eigenvalue representation is more significant for the prediction of $G$.

While the Adam optimizer is typically behind AdaBelief, it remains a valuable tool since it can help us understand whether a given molecular descriptor presents more learning-ready information for the prediction of a given output property. Such findings can be used for engineering larger-scale experiments, with more descriptors and molecules.

The classification approach presented in this paper can provide information of the mathematical or physicochemical kind regarding the selected training optimizer, inputs, outputs, and the relations between one another. This insight can be used to guide and hone further experiments showing more ambition. For example, a different dataset could be used, such as PC9 \cite{Glavatskikh2019,Cauchy2019} given its greater chemical diversity while remaining of similar size to QM9 \cite{Ramakrishnan2014}; the GDB family \cite{Fink2005,Fink2007,Blum2009,Ruddigkeit2012} is another interesting option, which lists molecules with up to 17 heavy atoms.

Alternatively, experiments with a higher precision goal could be undertaken, perhaps now using narrower intervals of interest, or using a greater amount of bins, thus resulting in a greater amount of classes while maintaining the relative similarity between the cardinality of each class. This applies to $H$, $G$, and ZPVE, given their positive performance for our experiments with 25 bins and 5 or 7 classes. Regression experiments could also be explored.

For $\Cv$, on the other hand, prediction quality must be refined before more granular experiments are to be attempted. Let us consider the low- and high-resolution descriptors we used: the main difference between them is the explicit encoding of interatomic distances, and its inclusion had a positive effect in the prediction quality. This is consistent with Einstein's and Debye's statistical-mechanics models for $\Cv$ of solids \cite{Einstein1907,Debye1912}, where bond strength is an important factor.

Therefore, desirable representations for future testing would include bond information. Examples of such descriptors are Bag of Bonds \cite{Hansen2015}, bonds-and-angles machine learning (BAML) \cite{Huang2016}, histograms of distances, angles, and dihedral angles \cite{Faber2017}, Behler-Parrinello or Smith symmetry functions \cite{Behler2007,Behler2011,Smith2017-1}. With the performance baseline established in the aforementioned experiments, comparisons involving new descriptors are attainable.

%
%
%
%

\section*{Acknowledgements}


The authors thank Dr. Luis Alarc\'on-Ramos for his very kind help on hardware-related matters, as well as prompt technical assistance.

The authors thank Valentin Bogad for very swift and in-depth
assistance on diverse matters pertaining implementation and regarding the \texttt{julia} programming language.

L. A.-T. acknowledges support from scholarship \texttt{2020-000013-01NACF-11160} from CONACyT.

L. A.-T. thanks Dr. M. Davar for persistent discussion and optimization on diverse writing-adjacent aspects.

\printbibliography[heading=bibintoc,title={Bibliography}]

\end{document}